\documentclass{emulateapj}

\newcommand{\hei}{He~{\sc i} 2.112 $\mu m$}
\newcommand{\brg}{Br$\gamma$}
\newcommand{\mum}{\ifmmode \mu m \else $\mu m$\fi}

\newcommand{\teff}{\ifmmode T_{\rm eff} \else $T_{\mathrm{eff}}$\fi}
\newcommand{\logg}{\ifmmode \log g \else $\log g$\fi}
\newcommand{\lL}{\ifmmode \log \frac{L}{L_{\odot}} \else $\log \frac{L}{L_{\odot}}$\fi}

\newcommand{\kms}{km s$^{-1}$}
\newcommand{\msun}{\ifmmode M_{\odot} \else M$_{\odot}$\fi}
\newcommand{\zsun}{\ifmmode Z_{\odot} \else Z$_{\odot}$\fi}
\newcommand{\lsun}{\ifmmode L_{\odot} \else L$_{\odot}$\fi}
\newcommand{\rsun}{\ifmmode R_{\odot} \else R$_{\odot}$\fi}

\slugcomment{}

\shorttitle{IRS16SW: a massive binary in the GC}
\shortauthors{Martins et al.}

\begin{document}

\title{GCIRS16SW: a massive eclipsing binary in the Galactic Center \footnote{Based on observations collected at the ESO Very Large Telescope (programs 075.B-0547 and 076.B-0259)}}

\author{F. Martins \altaffilmark{1}, S. Trippe \altaffilmark{1}, T. Paumard \altaffilmark{1}, T. Ott \altaffilmark{1}, R. Genzel \altaffilmark{1,2}, G. Rauw \altaffilmark{3}, F. Eisenhauer \altaffilmark{1}, S. Gillessen \altaffilmark{1}, H. Maness \altaffilmark{4}, R. Abuter \altaffilmark{5}}

\email{martins@mpe.mpg.de}

\altaffiltext{1}{MPE, Postfach 1312, D-85741, Garching, Germany}
\altaffiltext{2}{Department of Physics, University of California, CA 94720, Berkeley, USA}
\altaffiltext{3}{Institut d'Astrophysique et de G\'eophysique, Universit\'e de Li\`ege, All\'ee du 6 Ao\^ut 17, B\^at. B5c, 4000 Li\`ege, Belgium}
\altaffiltext{2}{Department of Astronomy, University of California, CA 94720, Berkeley, USA}
\altaffiltext{5}{ESO, Karl-Schwarzschild-Str. 2, D-85748 Garching, Germany}

\begin{abstract}
We report on the spectroscopic monitoring of GCIRS16SW, an Ofpe/WN9
star and LBV candidate in the central parsec of the Galaxy. SINFONI
observations show strong daily spectroscopic changes in the K
band. Radial velocities are derived from the He~{\sc i} $2.112 \mu m$
line complex and vary regularly with a period of 19.45 days,
indicating that the star is most likely an eclipsing binary. Under
various assumptions, we are able to derive a mass of $\sim$ 50 \msun\
for each component.
\end{abstract}



\section{Introduction}
\label{intro}

The central cluster constitutes one of the largest concentrations of
massive stars in the Galaxy \citep{genzel03}. Nearly 100 OB and
Wolf-Rayet stars are confined in a compact region of radius $\sim$ 0.5
parsec centered on the super-massive black hole associated with the
radio source SgrA* \citep{paumard06}. Among this population of young
massive stars, six are thought to be Luminous Blue Variables
(LBV): IRS16NE, IRS16C, IRS16NW, IRS16SW, IRS33E and IRS34W
\citep{paumard04,trippe06}. LBVs are evolved massive stars
experiencing strong variability in both photometry and spectroscopy
due to their proximity to the Humphreys-Davidson limit \citep{hd94}, a
region of the HR diagram where the luminosity of the stars reaches the
Eddington luminosity so that instabilities develop in their atmospheres,
leading to strong mass ejection and drastic changes in the stellar
properties (\teff, radius).
The six stars mentioned above are only LBV ``candidates'' (LBVc)
since they have not been observed to experience the strong outbursts
and photometric changes typical of bona fide LBVs such as $\eta$ Car \citep{dh97}. However, their luminosities and
spectra are very similar to stars known to be ``quiescent'' LBV,
i.e. stars having experienced an LBV event in the past and being now
in a more stable phase. In addition, one of them - IRS34W - has shown
photometric variability on timescales of months to years which was
interpreted as the formation of dust from material previously ejected
by an LBV outburst \citep{trippe06}.

Among these six stars, IRS16SW deserves special attention. This star was
claimed to be a massive eclipsing binary by \citet{ott99} since its K
band magnitude displays regular variations with a periodicity of 9.72
days. However, the absence of a second eclipse in the light-curve lead
\citet{depoy04} to the conclusion that the binary scenario was not
correct, and that IRS16SW was instead a pulsating massive star, a
class of star predicted by theory but not observed so far.

Here, we present results of the spectroscopic monitoring of IRS16SW
revealing periodic variations of radial velocities which are
interpreted as the signature of a massive spectroscopic and eclipsing
binary .

\begin{figure*}
\epsscale{0.9}
\begin{center}
\includegraphics[width=8.5cm,height=6.5cm]{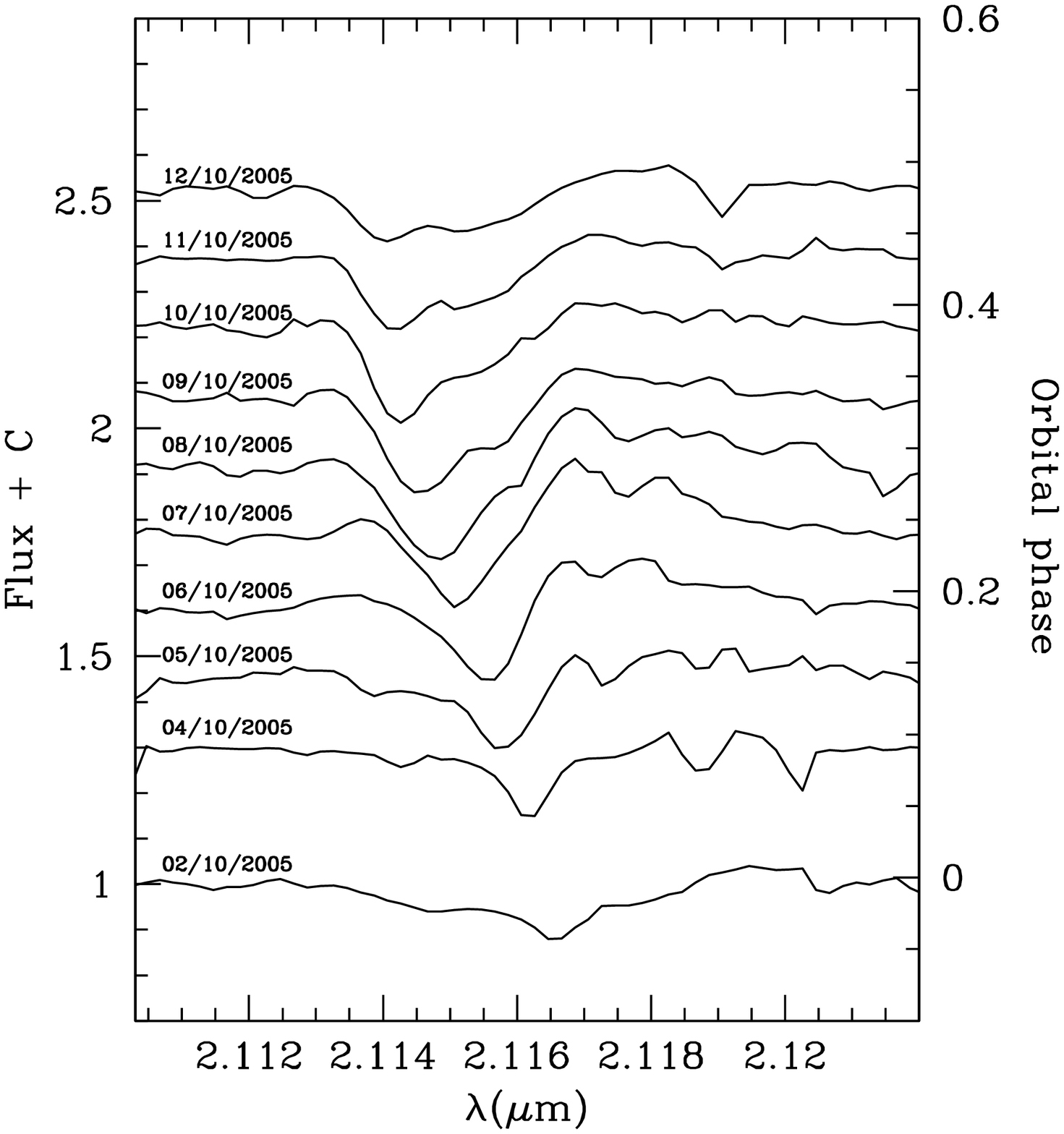}
\includegraphics[width=8.5cm,height=6.5cm]{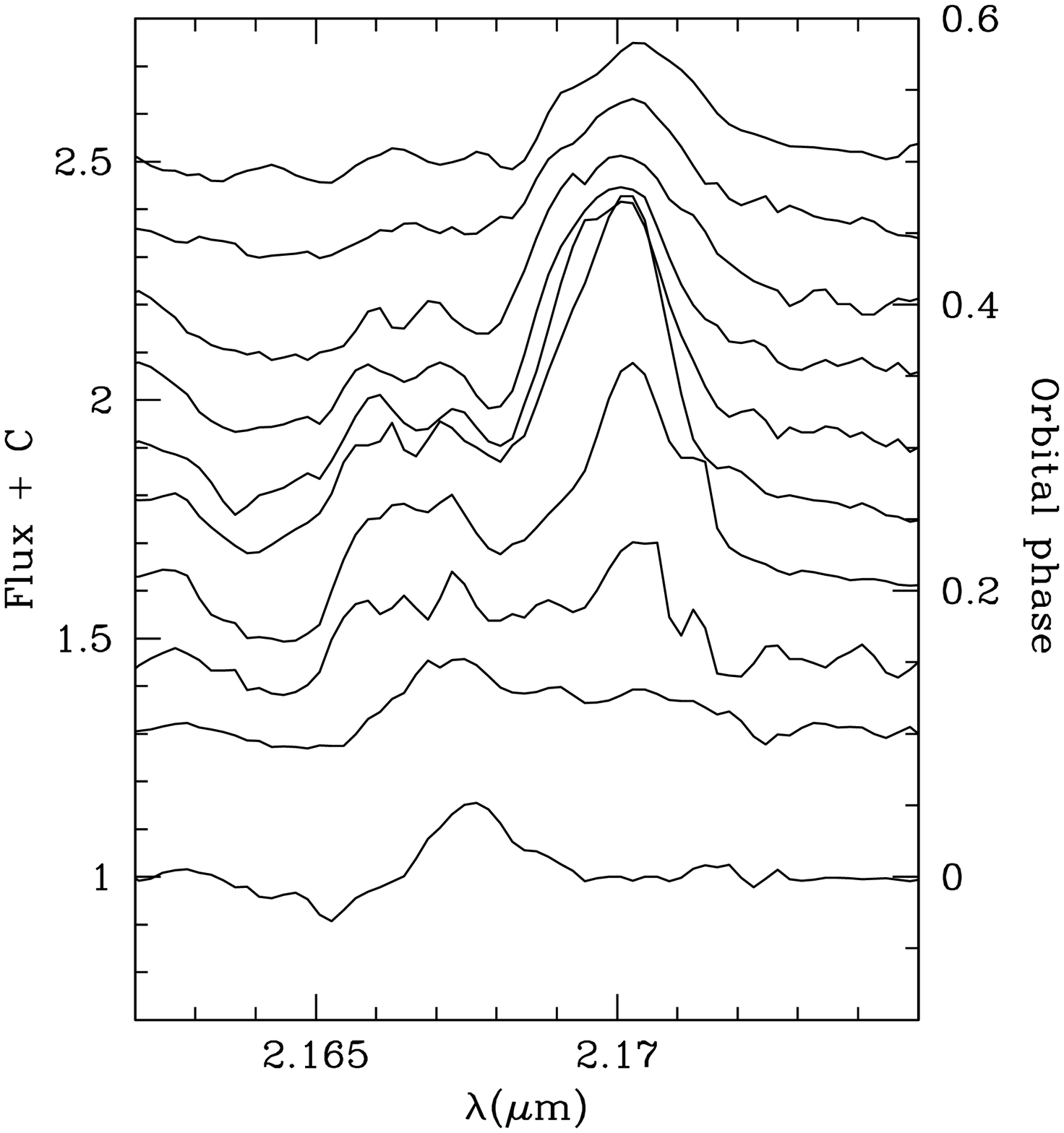}
\end{center}
\caption{Montage of spectra of IRS16SW around the region of He~{\sc i}
  2.112 $\mu m$ (left) and \brg\ (right) taken between October
  2$^{nd}$ and October 12$^{th}$ 2005. The changes in both the line
  shape and position are clearly seen. \label{var_lines}}
\end{figure*}


\section{Observations and data reduction}
\label{observ}

We used SINFONI \citep{frank03} on the ESO/VLT to obtain spectra of
IRS16SW.  Observations were carried out under seeing limited
conditions and were performed on August 28$^{\rm th}$ - September
1$^{\rm st}$, September 4$^{\rm th}$, October 2$^{\rm nd}$, and
October 4$^{\rm th}$ - 12$^{\rm th}$ 2005, and March 18$^{\rm th}$ -
21$^{\rm st}$ 2006. In order to get the best spectral resolution
available with SINFONI (R = 4000) we restricted ourselves to the K
band. Short exposures ($2 \times 60$ seconds) were sufficient to
obtain S/N $\sim$ 30. Data reduction was performed as in
\citet{frank05}. The final spectra were subsequently carefully
extracted from the ``data cubes'' by selecting a circular aperture
(radius of 3 pixels) centered on the star and by subtracting from it
an annulus of inner (outer) radius 3 (4) pixels. This procedure
allowed a good removal of nebular contamination.


\section{Results}
\label{results}

\subsection{Spectroscopic variability and radial velocities}
\label{var_rv}

Fig.\ \ref{var_lines} shows the variation of the \hei\ and \brg\ lines
with time. It is obvious that not only the line shape
but also the position of the centroid varies.  In order to test
the binary scenario, we have derived radial velocities (RV). For that
purpose, we have used the line at 2.112 $\mu m$ since it is less
affected by wind emission than other lines and is formed closer to the
photosphere, allowing a better estimate of the star's motion. This
line is however a blend of at least two He lines, and synthetic
spectra computed with atmosphere models reveal that the position of
the strongest absorption part of the profile can vary by several 100
\kms\ around 2.112 $\mu m$ depending on the stellar and wind
properties.  We have thus adopted this wavelength as our reference,
but we stress that the absolute value of the RV may be systematically
shifted compared to the real value due to this choice.  In practice,
we have measured the position of the maximum absorption trough in the
line complex and computed the radial velocity from the wavelength
shift compared to the adopted reference wavelength. When the line
shows a double peak, we have always measured the position of the
deepest absorption part of the profile (which also turned out to
always be the bluest). This implicitely assumes that if the star is a
binary, this absorption is always produced by the same star.  Note
that when present, the two absorption peaks are separated by the
theoretical spacing between the two Helium lines around 2.112 $\mu m$:
170 \kms.  Hence we conclude that the second peak most likely
comes from the same star. 
The derived
RV are presented in Fig.\ \ref{fit_vrad} and follow very nicely an almost
exact cosine curve (see below). This is a strong indication that
IRS16SW is a binary star (and justifies a posteriori our method to
derive RVs).

A period folding analysis applied to the RV curve gives 19.3 $\pm$ 0.4
days, in good agreement with the K band light-curve
analysis. \citet{depoy04} derived a period of 9.725 $\pm 0.005$ days
but argue that if this light-curve was to be produced by a binary
star, the absence of second minimum should point to a system composed
of two stars with the same K band luminosity, and consequently to a
true period of $2 \times 9.725 = 19.45$ days. A re-analysis of the
photometric data of \citet{ott99} in view of these new results
confirms that a period of 19.447 $\pm 0.011$ days is indeed present in
the period folding diagram \citep[for a description of the method,
see][]{ott99}. We conclude that IRS16SW is most likely a single line
spectroscopic (SB1) and eclipsing binary with an orbital period of
19.45 days.

\subsection{Orbital solution and physical parameters}
\label{orb_sol}

\begin{figure}
\epsscale{0.9}
\begin{center}
\includegraphics[width=8.5cm,height=6.5cm]{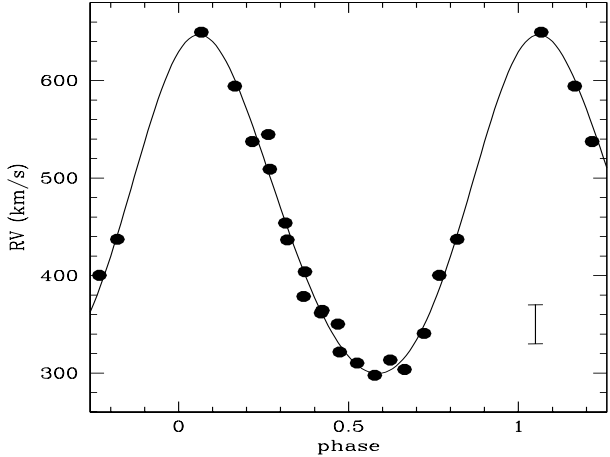}
\end{center}
\caption{Radial velocity curve of IRS16SW together with the best
orbital solution (for P = 19.45 days, solid line). The typical
uncertainty on the radial velocity is $\pm$ 20 \kms. Parameters for
the best fit solution are given in Table
\ref{tab_param_orbit}. \label{fit_vrad}}
\end{figure}

\begin{figure}
\epsscale{0.9}
\begin{center}
\includegraphics[width=8.5cm,height=6.5cm]{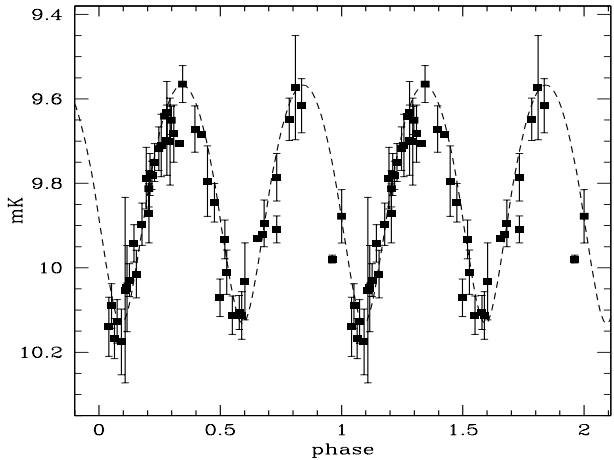}
\end{center}
\caption{K band light-curve displayed for a period of 19.45 days,
together with the best fit (dashed line). Data are from
\citet{ott99}. The two eclipses are similar, indicating that both
components of the binary have the same K band magnitude. Note also the
absence of plateau which suggests that contact is
achieved. \label{fit_kcurve}}
\end{figure}

With the RVs in hand, we have performed an orbital
solution for an SB1 binary using the method of \citet{rauw00}, which
is based on the Wolfe, Horak \& Storer algorithm \citep{wolfe67}. The
resulting parameters are given in Table \ref{tab_param_orbit}. Note
the small eccentricity justifying that the fit of the RV curve in
Fig.\ \ref{fit_vrad} is almost a cosine. The orbital solution also
gives the mass function

\begin{equation}
f(m) = \frac{(M_{2} \sin i)^{3}}{(M_{1}+M_{2})^{2}} = 10.47 M_{\odot}
\label{eq_fm}
\end{equation}

\noindent from which one can estimate the individual masses of each
component ($M_{1}$ and $M_{2}$) if one uses in addition
Kepler's third law

\begin{equation}
M_{1}+M_{2} = \frac{4 \pi^{2} r^{3}}{G P^{2}}
\label{kepler3}
\end{equation}

\noindent where $r$ is the separation between the two stars and $P$
the period. However, one needs an estimate of 1) the inclination and
2) the separation. As for the latter, the absence of plateau in the K
band light-curve indicates that contact is achieved in the binary: as
soon as the primary eclipse ends, the secondary eclipse starts. Hence,
one can assume that $r$ is simply the sum of the radii of
the two components. The value of $a \sin i$ we find (and $a$ itself,
$\sin i$ being close to 1, see below) is similar to the stellar radius
of LBVc stars derived by \citet{paco97}, so we assume $r = 2 \times
a$.  In that configuration, the two stars, which we know have similar
K magnitude, have similar radii and rotate around each other, the
center of mass being the contact point. To get an estimate of the
inclination, we have used the software NIGHTFALL \footnote{software
developed by R. Wichmann and freely available at the following URL
http://www.hs.uni-hamburg.de/DE/Ins/Per/Wichmann/Nightfall.html} to
fit the light-curve (see result in Fig.\ \ref{fit_kcurve}). Assuming a
similar effective temperature of 28000 K for both components
\citep[see][]{paco97} and a light ratio of one, we
obtained a reasonable fit with an inclination $i \sim 70 \deg$. Note
that we had to adopt Roche lobe filling factors of 1.3 (the maximum
allowed value in NIGHTFALL) to correctly reproduce the light-curve. If
filling factors lower than one are used, the light-curve can not be
reproduced correctly as the ``peaks'' are too broad \citep[a problem
encountered by][]{depoy04}. This confirms that contact is achieved and
justifies our assumption that the separation is the sum of the stellar
radii.

Given the limitations on the light-curve fit and the complications in
the physics of the star due to contact (mass transfer, departure from
sphericity due to gravitational interaction, hot spot in interaction
region), we stress that our estimate of the inclination is only
indicative. It is however consistent with the value expected for an
eclipsing binary. With $i \sim 70 \deg$, and using Eq.\ \ref{eq_fm}
and \ref{kepler3}, we finally derive $M_{1} \sim M_{2} \sim 50
M_{\odot}$.

\begin{table}
\begin{center}
\caption{Orbital parameters - semi-major axis, eccentricity, systemic
velocity, amplitude, longitude of periastron - as derived from the
analysis of the radial velocity curve. \label{tab_param_orbit}}
\begin{tabular}{l|ll}
\tableline
$a sini$ [\rsun]  &         66.4 $\pm$ 2.2     &\\
$e$       &         0.088 $\pm$ 0.023          &\\
$v_{0}$ [\kms]  &         459.5 $\pm$ 3.6      &\\
$K1$ [\kms]     &         173.8 $\pm$ 5.5      &\\
$\omega$ [$\deg$] &         334.0 $\pm$ 18.3   &\\
\tableline
\end{tabular}
\end{center}
\end{table}

\subsection{Spectral disentangling}
\label{spec_disentang}

In order to get more insight into the properties of the components of
IRS16SW, we have attempted to disentangle the spectra using the method
of \citet{gl06}. In practice, the RVs of the primary are used to
evaluate an average spectrum in the primary's rest frame. This
provides an approximation of the primary spectrum which is then
shifted back into the observer's frame and subtracted from the
observed spectrum.
The residual is composed of the
secondary spectrum from which RVs can be estimated. The whole
procedure is then re-started inverting the role of the two components,
and is iterated until convergence.

\begin{figure}[t]
\epsscale{0.9}
\begin{center}
\includegraphics[width=8.5cm,height=6.5cm]{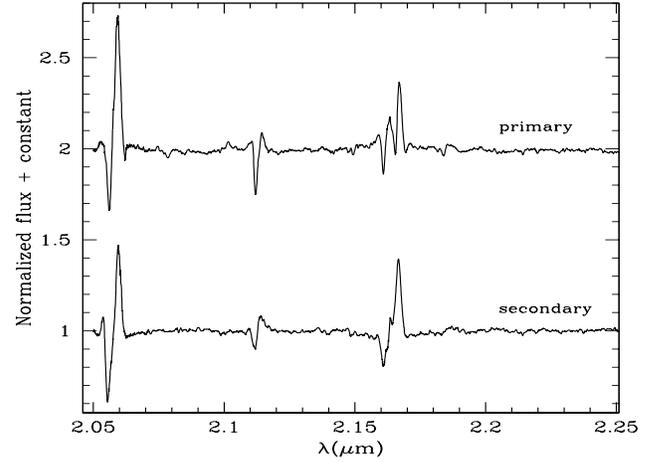}
\end{center}
\caption{Spectra of the primary (top) and secondary (bottom)
components of IRS16SW as obtained from the spectral disentangling
analysis. These spectra should be
interpreted with caution since they were obtained under several
assumptions (see text). \label{fig_disent}}
\end{figure}

Due to the limited number of spectral lines and S/N ratio of our
spectra, no solution could be found leaving all parameters free. We
therefore decided to freeze the primary RVs at their values determined
above. Given the results of the orbital solution, we also decided to
set the mass ratio to 1.0, and assumed that both components have the
same systemic velocities. In that case, convergence could be achieved
and the resulting spectra are shown in Fig.\ \ref{fig_disent}.  These
spectra should be interpreted with caution. Not only they were
obtained under the assumption that the mass ratio is 1.0, but the
procedure used also implies that the spectra are free of contamination
by wind-wind collision or any other interaction in the contact
region. With these restrictions in mind, the main qualitative
conclusion we draw is that the spectra of both stars, and consequently
their properties, are similar and typical of LBV candidates such as
the Pistol star \citep{figer99}.

\section{Discussion}
\label{disc}

\subsection{Binary versus pulsating variable}
\label{disc_bin}

\citet{depoy04} argue that IRS16SW was a pulsating massive star based
on the absence of second minimum in the light-curve and the difficulty
to fit this light-curve in the binary scenario (but see Sect.\
\ref{orb_sol}). They compare the observed variation in K magnitude to
the predictions of the dynamical models of \citet{dg00} and conclude
that there is a reasonable qualitative agreement. However, there are
some quantitative discrepancies. First, the amplitude of the variation
is much larger than predicted: although \citet{dg00} do not compute K
band photometry, one can estimate the variation in this band to be at
most 0.2 mag (inspection of their Table 2 reveals that the amplitude
of photometric variations decreases with wavelength and is $\lesssim
0.2$ mag in the I band), while we observe 0.55 mag.  Second, the
period we derive $-$ 19.45 days $-$ is larger than expected in the
pulsating scenario \citep[see Table 1 of][]{dg00}.

Concerning spectroscopic changes, although in principle one can not
completely rule out the possibility that they are due to motions of
the atmosphere and fluctuations of the physical parameters (\teff,
radius) due to pulsations \citep{dg00}, the timescales are again not
consistent: $\sim$ 1 day for pulsations compared to 19.45 days
observed. Besides, so far there are no theoretical predictions of
spectroscopic changes caused by pulsations to which we could compare
our observed spectra. The binary nature of IRS16SW is thus strongly
favored.

The absence of secondary eclipse in the light-curve is
explained by the similar K band magnitude of the two components.
This is another indication that both stars are very similar. 
\citet{depoy04} report the presence of a variation in $H-K$ on a
period of 9.725 days, $H-K$ being bluer when the system is brighter. A
similar trend was observed in the optical photometry of the
massive contact binary V606 Centauri \citep{lorenz99}. We interpret
this as a sign of heating in the contact zone, making
the spectral energy distribution in this region bluer. Massive
binaries are indeed known to produce X-rays through colliding winds,
and we might expect the same kind of interaction and heating around
the contact region. Again, since this region is seen twice during an
orbital revolution, an observed period half the true orbital period is
naturally derived from the $H-K$ curve.

\subsection{Stellar evolution and the LBV phenomenon}
\label{disc_evol}

Whether or not all massive stars go through the LBV phase is still
under debate. \citet{langer94} and \citet{pasquali97} argue that this
is the case, while other observational \citep{crowther95} and
theoretical \citep{mm05} studies indicate that the most massive stars
($M \gtrsim 60 \msun$) may skip this phase. This is an important issue
since although short, the LBV phase is crucial in the mass loss
history, and consequently in the subsequent evolution, of massive
stars.  Recent studies by \citet{smith06} even claim that most of the
mass of hot stars is lost during the LBV phase.  Here, we provide an
accurate measurement of the present mass of a candidate LBV, confirming
that a star with an initial mass larger than 50 \msun\ (and likely
of the order 60-70 \msun) may become a LBV. This is a strong
constraint for evolutionary models.

Of course, one could argue that the star's evolution was affected by
binarity. However, inspection of the spectral morphology and physical
properties \citep[see][]{paco97} of IRS16SW and the other LBVc in the
Galactic Center shows similarities. Our monitoring of IRS16SW also
includes the other LBVc in the Galactic Center. Except for IRS16NE,
none of these stars showed any spectroscopic variation, ruling out the
possibility that they are close binaries. IRS16NE showed some RV
fluctuations \citep[see also][]{tanner06}, but so far we do not have
enough data point to sample an hypothetical RV curve. Hence, the
similarity between the spectrum of IRS16SW and those of the other
single LBV candidates leads us to the conclusion that binarity has not
(yet?) significantly affected the evolution of IRS16SW. Since the mass
of the two components is similar, one can speculate that IRS16SW is
composed of two stars initially equally massive that have so far
evolved in parallel in a detached system, without influencing each
other. They may have just entered the LBV phase during which contact
was achieved due to their respective expansion. This event probably
happened very recently (the LBV phase lasting $\sim 10^{5} yr$) so
that the general properties of both components have not yet been
affected by mass transfer and binary evolution. Such a scenario is
consistent with both stars displaying similar spectra (but see Sect.\
\ref{spec_disentang} for caution words).

Assuming that IRS16SW was not affected by binary evolution, the
properties of the LBV candidate in IRS16SW can thus be used to
constrain evolutionary models of single massive stars.  In their
recent models including rotation, \citet{mm05} stressed that the LBV
phase is not systematically reached above 45 \msun. Here, we have an
example for which it is the case (under the assumption that IRS16SW
will turn $-$ or has already turned in the past $-$ into a genuine
LBV).

\acknowledgments

We thank all the ESO staff for their help during observations.
FM acknowledges support from the Alexander von Humboldt Foundation.



\end{document}